\documentclass[11pt]{article}
\usepackage[a4paper,margin=1in]{geometry}
\usepackage{authblk}
\usepackage{amsmath,amssymb}
\usepackage[numbers,sort&compress]{natbib}
\usepackage[hidelinks]{hyperref}

\AtBeginDocument{%
  \providecommand\BibTeX{{%
    \normalfont B\kern-0.5em{\scshape i\kern-0.25em b}\kern-0.8em\TeX}}}

\usepackage{xcolor}
\usepackage{todonotes}
\usepackage{graphicx}
\usepackage{csquotes}
\usepackage{subcaption}  
\usepackage{soul}
\usepackage{booktabs}

\usepackage{tcolorbox}
\usepackage{mdframed}
\usepackage{fvextra}
\usepackage{multirow}
\usepackage{fontawesome5}
\usepackage{twemojis}
\tcbuselibrary{listingsutf8}

\usepackage{tikz}
\usetikzlibrary{arrows.meta,positioning,calc}
\providecommand{\Description}[1]{}

\begin{document}

\title{Sensed but Not Told: Hidden Physiological Adaptation Shapes LLM Persuasion}

\author{Xiaoyan Wei\thanks{\href{mailto:xiaoyan.wei@adelaide.edu.au}{xiaoyan.wei@adelaide.edu.au}}}
\author{Yutong Qu}
\author{Yutong Li\thanks{\href{mailto:a1948101@adelaide.edu.au}{a1948101@adelaide.edu.au}}}
\author{Haiyi Li\thanks{\href{mailto:a1949007@adelaide.edu.au}{a1949007@adelaide.edu.au}}}
\author{Wei Emma Zhang\thanks{\href{mailto:wei.e.zhang@adelaide.edu.au}{wei.e.zhang@adelaide.edu.au}}}
\author{Hsiang-Ting Chen}
\affil{School of Computer Science and Information Technology, Adelaide University, Australia}
\renewcommand{\Affilfont}{\small}
\date{}

\maketitle

\begin{abstract}
Wearable sensors, including those embedded in VR headsets, can provide LLM-based chatbots with physiological signals that users do not explicitly communicate.
We tested whether such hidden input changes persuasion.
In a $2\times2$ within-subjects study ($N=24$), a chatbot argued against participants’ self-selected opinions with or without turn-level heart-rate and pupil-diameter features; sensing was disclosed halfway through the study. Before disclosure, sensing increased perceived persuasiveness by 1.42 points. Across the study, sensing was associated with a 0.57-point greater shift toward the chatbot’s position and 3.20 times the odds of any movement toward it. Persuasiveness ratings were lower after disclosure, although disclosure was confounded with session order.
Dialogue logs suggested a more declarative interaction style when sensing was on: chatbot replies contained question marks less often, and conversations contained fewer turns.
These findings motivate making biosignal-driven adaptation inspectable and providing a genuinely non-sensed mode.
\end{abstract}


\section{Introduction}

LLM-based conversational agents are increasingly capable of changing peoples' minds: they match human writers on policy arguments~\cite{Bai2025LLMPersuation,palmer2023governance}, outperform humans in live debates when given basic information about their opponent~\cite{Salvi2025Persuade}, and can reduce conspiracy beliefs through short dialogues~\cite{costello2024reduceConspiracy}.
Personalization may extend beyond personal information, such as demographic characteristics, preferences, to real-time physiological data.
Headsets, watches, and earbuds can stream heart rate, pupil diameter, and related measures while a conversation unfolds.
An LLM could use these signals as additional context when adapting its arguments, without relying on what the user explicitly communicates.

Prior work shows that physiological adaptation can benefit users, for example by enabling coaching systems to respond to stress or fatigue~\cite{Spelt2022Physiology,idrees2024persuasive,wu2024mindshift,dongre2024physiology}. The same capability, however, raises concerns about manipulation when used covertly to increase persuasive influence~\cite{danry2024deceptive,liu2025llm,salehzadeh2024wearable}. Disclosure offers one safeguard, but it may affect how people evaluate a persuader without changing whether their attitudes shift~\cite{Lim2023Disclosure,bohm2023devalueLLM}.
We therefore examine perceived persuasiveness and attitude movement separately,  both before and after participants learn that the system can access physiological signals.

Studying these effects requires varying the model's access to physiological signals without visibly changing the sensing setup. We used sensors integrated into a VR headset that participants wore throughout the study. The physical setup remained unchanged whether or not heart-rate and pupil features were passed to the model.
Twenty-four participants each held four conversations with a GPT-4 Turbo chatbot arguing against a self-selected opinion. The $2\times2$ design crossed \emph{sensing} (on or off) with \emph{awareness} (before or after disclosure); in the post-disclosure sensing-off cell, participants believed they were being sensed although no signals were assigned to the model. We ask:

\begin{itemize}
\item \textbf{RQ1:} When physiological sensing is hidden, does access to the signals change perceived persuasiveness and the direction of immediate attitude movement?
\item \textbf{RQ2:} How do these outcomes differ after participants are told about the sensing?
\end{itemize}

While sensing was hidden, the adaptive chatbot was rated 1.42 points more persuasive on a seven-point scale. After accounting for which sensing condition participants encountered first, the estimated difference remained similar at 1.34 points.
Across the 93 conversations, sensing was associated with an additional 0.57-point shift toward the chatbot's position.
Attitude ratings shifted by at least one scale point toward the chatbot's position in 25 of 47 sensing-on conversations (53\%), compared with 12 of 46 sensing-off conversations (26\%).
After disclosure, perceived-persuasiveness ratings were lower, and the sensing-on advantage decreased from 1.42 to 0.33 points. Because disclosure always occurred at the session midpoint, these changes cannot be separated from session-order effects.
Dialogue logs showed fewer questions and shorter conversations when sensing was on, but did not explain why participants responded differently.

This paper contributes a focused demonstration that hidden physiological input can change both the experience and immediate direction of an LLM-mediated persuasive encounter.
The findings also suggest that disclosure of sensing alone does not make physiological adaptation transparent: ratings declined after participants were informed about sensing, while its influence on individual responses remained difficult to identify.
Because disclosure always occurred at the session midpoint rather than being independently randomized, we treat comparisons between the two blocks as secondary.

\section{Related Work}

\subsection{Personalised persuasion by LLMs}

Tailoring is a central strategy in persuasive technology~\cite{Fogg2002PersuasiveTechnology,kaptein2012heterogeneity,kaptein2015personalizing}.
LLMs make personalized messaging inexpensive and scalable.
Messages tailored to recipients' personalities or values can outperform generic alternatives~\cite{matz2024potential}, and GPT-4 has outperformed human debaters when given demographic information about its interlocutor~\cite{Salvi2025Persuade}.
However, other large studies find little added value from microtargeting over strong generic prompts~\cite{doi:10.1073/pnas.2403116121,hackenburg2025levers}.

Beyond explicit microtargeting, an agent's conversational style can itself shape persuasion. Recent CHI studies varied emotional language and linguistically expressed personality in charitable-giving conversations.
These manipulations changed trust and empathy yet without producing consistent corresponding changes in donation decisions~\cite{Genc2025Persuasion,genc2026bots}.
A recent study of deceptive LLM arguments likewise found that persuasive effects varied with the argument's relation to a participant's prior stance and with individual factors such as topic knowledge and involvement~\cite{yeo2026deception}.
Together, these works show that persuasive effects are heterogeneous and that evaluations of an agent need not align with behavioural or attitudinal outcomes.
We therefore treat perceived persuasiveness and attitude movement as distinct outcomes.

\subsection{Physiologically adaptive agents}
Unlike profile-based personalization, which is typically established before a conversation, biosignals such as heart rate and pupil diameter can provide continuous, passively sensed context as the interaction unfolds~\cite{schmidt2018introducing,liu2023biosignal,iriarte2021evaluating}. Physiological adaptation can therefore respond to changes within a person rather than only to differences between people.

Recent systems have supplied physiological and behavioural data to LLMs for personal-health exploration, empathic stress-management dialogue, and interventions against problematic smartphone use~\cite{fang2024physiollm,dongre2024physiology,wu2024mindshift}.
Multiple recent works~\cite{neupane2025wearable,wang2025healthguru} extended this direction with wearable-triggered stress support and sleep recommendations grounded in wearable and contextual data.
More recently, Liu et al. ~\cite{liu2026sensing} proposed a system combining electrodermal activity and mouse behaviour to time LLM-generated assistance during surveys.
These systems demonstrate the value of sensing for support selected or accepted by the user. Less is known about the same capability in persuasive dialogue when users have neither requested the adaptation nor been told how their signals are being used.
Our study examines this less visible use of physiological adaptation.

\subsection{Awareness, forewarning and resistance}

The Persuasion Knowledge Model suggests that people reinterpret an influence attempt once they recognise the tactic behind it~\cite{friestad1994persuasion,campbell2000consumers}. Forewarning can also prompt counter-arguing and resistance~\cite{petty1977forewarning,compton2013inoculation}. Disclosure may therefore change how people interpret both a message and its source, even when it does not change their response to the argument itself. Prior studies support this distinction: disclosure often changes evaluations of AI-generated content more readily than selection or sharing behaviour~\cite{Lim2023Disclosure,Bashardoust2024ShareGenAI,bohm2023devalueLLM}.

Recent works further show that disclosure effects depend on both context and outcome.
Gamage et al.~\cite{gamage2025labeling} shows that labels for synthetic social-media content changed beliefs about AI involvement and trust in the labels, but had little effect on engagement.
In LLM-modified news, disclosure modestly improved trustworthiness without reducing perceived fairness or credibility~\cite{khatiwada2026news}.
Physiological adaptation introduces an additional concern because the system is not only generating the message but also sensing the person receiving it. A disclosure may make that persuasive tactic salient without revealing when or how a particular response was adapted. We therefore examine whether hidden sensing changes the persuasive encounter and, secondarily, what changes after sensing is disclosed.

\section{Chatbot Design and Formative Evaluation}

To examine this gap, we developed a VR-based persuasive chatbot that could receive turn-level physiological features without any visible change to the physical device setup.
Before the main user study, we conducted a formative evaluation to select a signal representation and verify that the adaptive configuration produced observable changes in the chatbot's responses. This section describes the chatbot and formative evaluation; Section~\ref{sec:study} then evaluates how the configuration affected participants' persuasive encounters.

\textbf{Sensing.} Participants wore an HP Reverb G2 Omnicept Edition headset. Its visor contains a PPG heart-rate sensor and eye cameras reporting pupil diameter at 120~Hz; the SDK exposes heart rate at 1~Hz. Because the sensors are built in, the equipment setup is identical whether or not the physiological data reached the chatbot, which enables the design of a hidden sensing condition.

\textbf{Physiological Features.} For each participant response turn, we defined a sensing window that began when the chatbot's message appeared and ended when the participant submitted their reply.
We supplied the turn standard deviation of heart rate and left and right pupil diameter, together with resting means. The 1~Hz heart-rate stream supports only coarse turn-level context, not HRV analysis. We treat these features as ambiguous contextual indicators, not direct measurements of emotion~\cite{kreibig2010autonomic,levenson2014autonomic}.

\textbf{Prompting.} GPT-4 Turbo was assigned the position opposite to the participant and instructed to argue in at most two sentences (Supplementary Appendix).
In sensing-on conversations, the model received both the features above and instructions to infer the participant's state, choose a tone, and adapt without mentioning the use of signals in the conversation.
In sensing-off conversations, both the data and the state-inference instructions were removed.

\textbf{Formative evaluation.} Five participants supplied 63 utterances paired with elevated heart rate and a self-reported emotion. Four HCI researchers, blind to condition, compared responses generated from five signal representations. Current-window standard deviation received 153 of 252 best-match votes (60.7\%), more than the alternatives combined. We then crossed signal presence with an instruction to use it. Judges identified emotion-based adaptation in 251 of 252 judgements (99.6\%) when both were present, versus 38 of 252 (15.1\%) for instruction alone, 7 of 252 (2.8\%) for data alone, and 18 of 252 (7.1\%) for neither. This establishes observable response differences, not accurate emotion inference; full counts appear in the Supplementary Appendix.

\section{User Study}
\label{sec:study}

\subsection{Design}

\begin{figure}
    \centering
    \includegraphics[width=\linewidth]{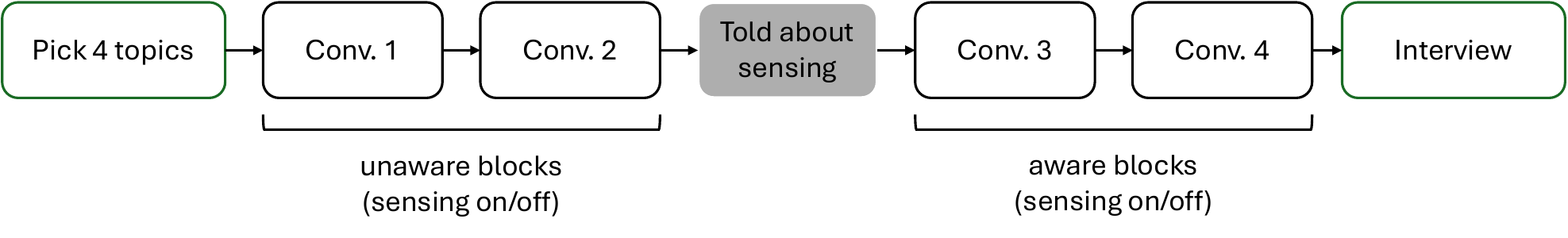}
    \caption{Study procedure. Each participant held four persuasive conversations, one per cell. The two unaware conversations always came first, because awareness cannot be withdrawn; the order of sensing on and off within each block varied between participants.}
    \Description{A timeline shows topic selection, two conversations before disclosure, disclosure of physiological sensing, two conversations after disclosure, and a final interview. Sensing is on for one conversation and off for the other within each block.}
    \label{fig:design}
\end{figure}

The study used a $2\times2$ within-participants design crossing \textbf{Sensing} and \textbf{Awareness} . Sensing determined whether turn-level heart-rate and pupil-diameter features were passed to the model. Awareness was operationalized by whether a conversation occurred before or after participants were told that the chatbot could access these signals.
This produced four cells: (\textbf{unaware}, \textbf{sensing on}), (\textbf{unaware}, \textbf{sensing off}), (\textbf{aware}, \textbf{sensing on}), and (\textbf{aware}, \textbf{sensing off}).
The  (\textbf{aware}, \textbf{sensing off}) cell provided an expectancy comparison: participants had been told about the chatbot's sensing capability, but no physiological features were passed to the model in that conversation.

As shown in Fig.~\ref{fig:design}, each participant first completed (\textbf{unaware}, \textbf{sensing on}), (\textbf{unaware}, \textbf{sensing off}) conditions (in counter-balanced order among participants).
We use this pre-disclosure, within-participant contrast as our primary comparison.
Participants were then told about the chatbot's access to their physiological signals and completed the  (\textbf{aware}, \textbf{sensing on}), and (\textbf{aware}, \textbf{sensing off}) conditions (again in counter-balanced order).

We fixed the block order because participants could not return to an unaware state after disclosure~\cite{poulton1982influential}. Disclosure could also make the sensing more salient or heighten participants' sense of being observed~\cite{mccambridge2014hawthorne}. Because the post-disclosure block always occurred later, differences between blocks may reflect disclosure, session order, or both. We therefore do not interpret between-block differences as causal effects of awareness; our primary conclusions concern the pre-disclosure sensing contrast.

\textbf{Ethical considerations.} The unaware conditions involved withholding the fact that the chatbot received heart-rate and pupil data during those conversations. The approved protocol limited this deception to the first two conversations and included a final debrief.
The topics concerned everyday opinions rather than personalized health, financial, or safety decisions, and participants could stop a conversation at any time.

\subsection{Participants and apparatus}

We recruited 24 participants (14 women, 10 men; aged 18--40) through university channels. Seventeen conducted the conversations in Mandarin and seven in English. The protocol was approved by the university ethics committee; participants gave informed consent, were told about the sensing before the aware block, were debriefed after the study, and received a \$20 AUD gift card. All sessions used the headset and chatbot described above.

\subsection{Interface}
\label{sec:interface}

\begin{figure}
    \centering
    \includegraphics[width=0.85\linewidth]{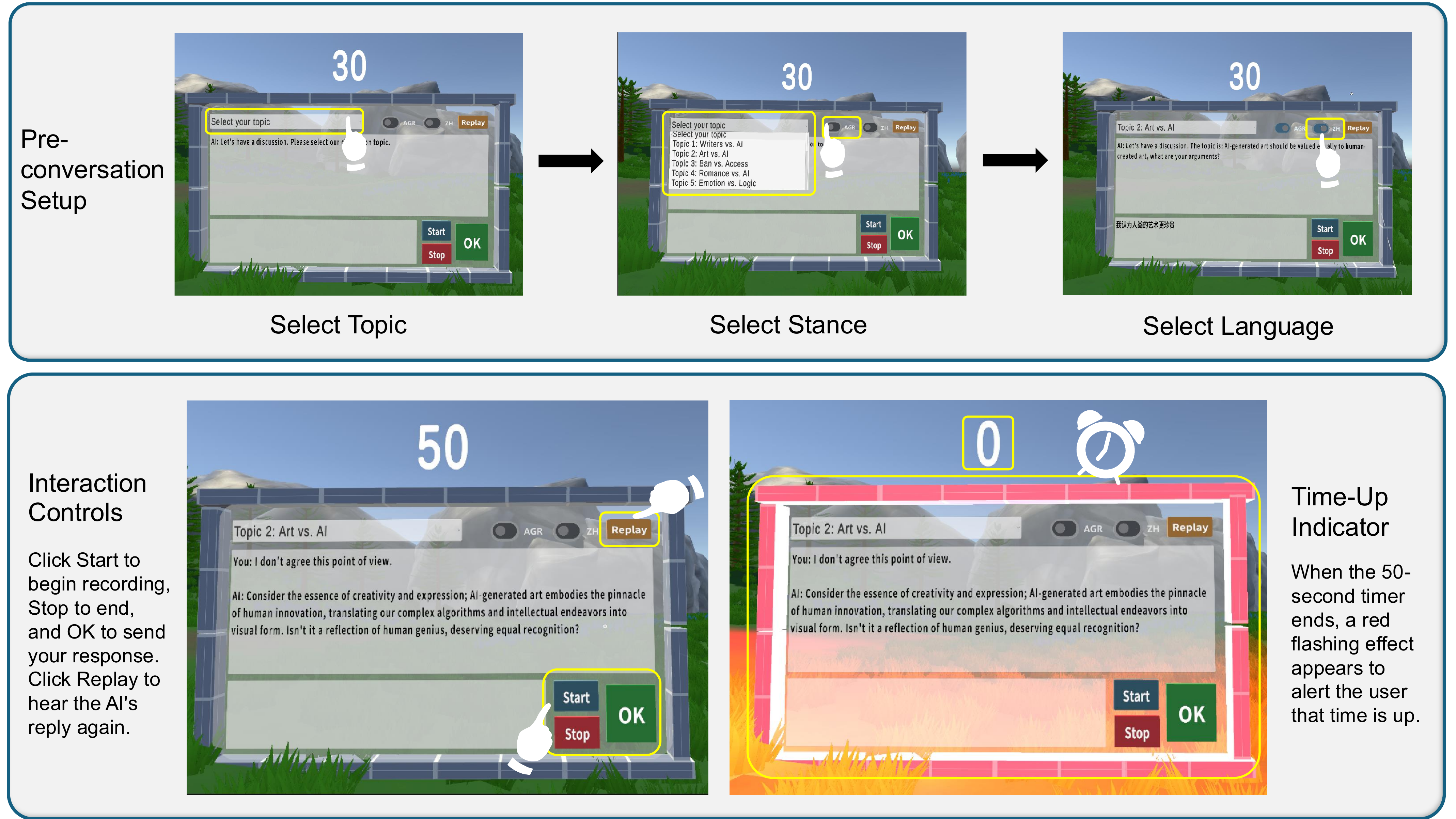}
    \caption{The VR conversation interface. Top: topic, stance and language selection. Bottom: voice controls and the timeout indicator.}
    \Description{Screenshots of the VR interface show the topic, stance, and language menus, followed by the conversation screen with start, stop, replay, and submit controls and a visible countdown timer.}
    \label{fig:interface}
\end{figure}

Conversations took place in a plain VR environment to reduce confound (Fig.~\ref{fig:interface}). All input was spoken: \emph{Start} and \emph{Stop} controlled recording, \emph{OK} sent the utterance, and \emph{Replay} repeated the chatbot's last message. We avoided text entry, which is slow and disruptive in VR~\cite{grubert2018text,dube2019text}, and avoided an avatar, since embodied agents invite social attributions that would add noise to judgements of the conversation itself~\cite{heidicker2017influence,zanbaka2006cat}.

\subsection{Procedure}

Fig.~\ref{fig:design} shows the experimental procedure.
After a five-minute briefing on the VR setup and study workflow, participants rated their agreement with 20 statements on a 1--7 scale (Supplementary Appendix). The statements covered AI ethics, education, cost of living, and social media.
Each participant selected four topics that they held a clear position and the chatbot was instructed to argue the opposing view.

Each conversation lasted 6--8 minutes. The participant opened by stating their view, the chatbot took the opposite side, and the two exchanged turns. To keep participants engaged and to elicit genuine reactions, they had 50 seconds to reply to each message, indicated by a flashing red light; after three timeouts the session ended and a different topic was chosen.
After each conversation they re-rated the topic and completed the post-conversation measures, then took a short break.
Before the third conversation, the experimenter told them that the chatbot would now read their heart rate and pupil size from the headset.
After all four conversations, a semi-structured interview (Supplementary Appendix) covered whether the conversation felt natural, whether they felt persuaded and why, what strategies they noticed, and whether they thought the chatbot was using non-verbal information.

\subsection{Measures}

Our primary outcome was a single \emph{perceived persuasiveness} rating on a 1--7 scale.
For each conversation, we calculated the pre–post attitude change and oriented its sign toward the position argued by the chatbot. Positive values were classified as movement toward the chatbot, zero as no change, and negative values as movement away from it.
We excluded three conversations that began at the neutral midpoint because they had no defined opposing direction.

Secondary measures were PANAS-SF positive and negative affect~\cite{watson1988development} and SAM valence, arousal, and dominance~\cite{bradley1994measuring}. We also collected the BFI-10 and scales on attitudes toward AI, social presence, and self-consciousness; these are outside the focused analysis. The Supplementary Appendix reports the secondary affect results and, for completeness, absolute attitude change. Interview and dialogue excerpts are illustrative rather than a formal qualitative analysis.

\subsection{Analysis}

For perceived persuasiveness, our primary analysis compared sensing-on and sensing-off ratings within the pre-disclosure block using a paired $t$-test. We report the mean difference, 95\% confidence interval, and standardized effect ($d_z$). Because the sensing order within this block was unbalanced, we also report an order-adjusted crossover estimate. A $2\times2$ repeated-measures ANOVA provides the sensing and block main effects and their interaction. We applied Holm correction across these three factorial tests and, separately, across the two within-block sensing comparisons.

For attitude movement, We used a Gaussian generalized estimating equation (GEE) model with participant-level clustering to account for the four conversations contributed by each participant. The outcome was the signed pre--post rating change for the 93 conversations with a defined direction. We first tested whether the association between sensing and attitude movement differed between the pre- and post-disclosure blocks. Because there was no evidence of such a difference, we estimated the average association with sensing across both blocks.
We then refitted this model with initial attitude strength and conversation position as additional covariates to check whether the estimate was robust to initial commitment and session order.
Finally, we repeated the analysis using a logistic GEE that compared movement toward the chatbot with either no change or movement away from it.

Affect outcomes were analyzed using the same $2\times2$ repeated-measures ANOVA without multiplicity adjustment and are interpreted descriptively. Because disclosure always occurred at the session midpoint, we also modeled perceived persuasiveness and negative affect using sensing condition, linear conversation position, and a post-disclosure step. This analysis distinguishes a gradual session trend from a change at the block boundary, but cannot identify a causal effect of disclosure.

Finally, we audited all 96 dialogue files, comprising 622 chatbot turns and 619 parseable replies. By sensing condition, we compared the number of turns per conversation, mean reply length, and the proportion of replies containing a question mark. We also tabulated the model-generated state and tone labels in sensing-on conversations. These exploratory measures identify observable differences in dialogue form; they do not test emotion-recognition accuracy or explain how sensing affected persuasion.

\section{Results}

All 96 conversations yielded complete rating and affect data; directional analyses used 93 after excluding three neutral starts.

\begin{figure}[t]
  \centering
  \includegraphics[width=\linewidth]{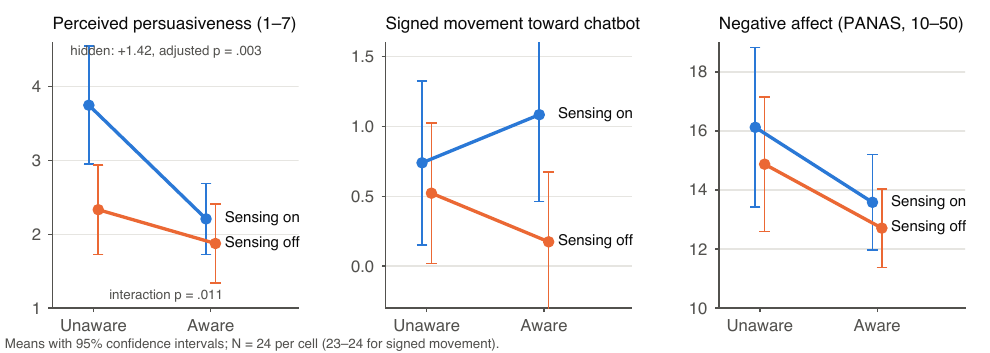}
  \caption{Perceived persuasiveness, signed attitude movement toward the chatbot, and negative affect by cell. The clearest interaction is in perceived persuasiveness: sensing produced a large advantage while hidden and a smaller difference after disclosure. Error bars are 95\% confidence intervals.}
  \Description{Three interaction plots compare sensing on and off before and after disclosure. Perceived persuasiveness is highest for hidden sensing. Signed attitude movement favours sensing overall. Negative affect is lower later in the session.}
  \label{fig:results}
\end{figure}

\subsection{Perceived persuasiveness}

Before disclosure, sensing raised perceived persuasiveness by 1.42 points, 95\% CI [0.66, 2.17], from $M=2.33$ ($SD=1.43$) to $M=3.75$ ($SD=1.89$), $t(23)=3.87$, Holm-adjusted $p=.003$, $d_z=.79$. The order-adjusted crossover estimate was similar: 1.34 points, 95\% CI [0.49, 2.18], $p=.004$, with no within-block drift ($p=.64$; Table~\ref{tab:robustness}).

Across all four cells, ratings were higher with sensing on, $F(1,23)=14.76$, adjusted $p=.002$, $\eta_p^2=.39$, and lower in the post-disclosure block, $F(1,23)=12.55$, adjusted $p=.003$, $\eta_p^2=.35$. Their interaction was also significant, $F(1,23)=7.73$, adjusted $p=.011$, $\eta_p^2=.25$ (Fig.~\ref{fig:results}, left). After disclosure, the sensing difference narrowed to 0.33 points, 95\% CI [$-0.11$, 0.78], adjusted $p=.40$. Hidden sensing produced the standout evaluation; the later block is interpreted separately below.

\begin{table}[t]
\centering
\small
\caption{Robustness of the focal persuasiveness and directional-attitude results. Point estimates are scale points except where marked OR. The first $p$ value is Holm-adjusted within the simple-effect family; sensitivity-model values are unadjusted.}
\label{tab:robustness}
\begin{tabular}{@{}p{0.46\linewidth}ccc@{}}
\toprule
Analysis & Estimate & 95\% CI & $p$ \\
\midrule
Hidden paired persuasiveness & +1.42 & [0.66, 2.17] & .003 \\
Hidden order-adjusted crossover & +1.34 & [0.49, 2.18] & .004 \\
Signed attitude, clustered GEE & +0.57 & [0.06, 1.07] & .029 \\
Signed attitude + strength and position & +0.54 & [0.07, 1.01] & .026 \\
Any movement toward chatbot & 3.20 OR & [1.50, 6.80] & .003 \\
\bottomrule
\end{tabular}
\end{table}

\subsection{Direction of attitude movement}

\begin{figure}[t]
  \centering
  \includegraphics[width=\linewidth]{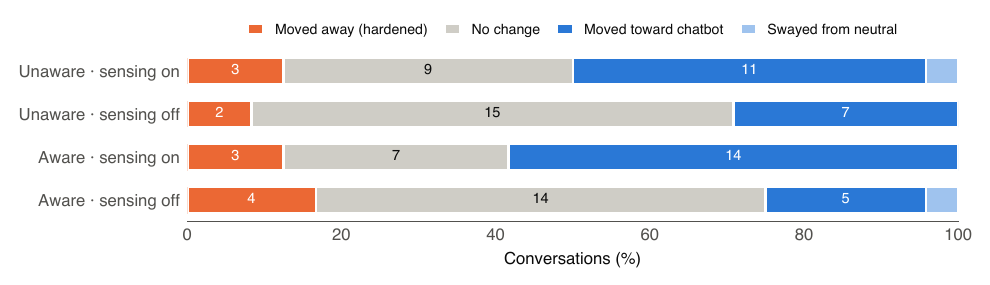}
  \caption{Direction of immediate attitude movement in each cell. Counts are printed within segments. Three conversations started from the neutral midpoint and have no defined opposing direction.}
  \Description{Stacked bars show more movement toward the chatbot when sensing is on. Counts toward the chatbot are 11 and 14 in the two sensing-on cells, compared with 7 and 5 in the sensing-off cells.}
  \label{fig:shift}
\end{figure}

The signed-movement interaction was not supported ($p=.20$). In the main-effects GEE, sensing was associated with 0.57 additional scale points toward the chatbot, 95\% CI [0.06, 1.07], $p=.029$; the post-disclosure block was not, $p=.99$. Adding starting-attitude strength and conversation position produced nearly the same estimate (0.54 points, $p=.026$).

The categorical result was consistent. Across 93 conversations, 37 moved toward the chatbot, 12 moved away, and 44 did not change. Movement occurred in 25 of 47 sensing-on conversations (53.2\%) versus 12 of 46 sensing-off conversations (26.1\%); hardening occurred six times in each condition (Fig.~\ref{fig:shift}). The interaction was not supported ($p=.27$); the reduced model estimated $OR=3.20$, 95\% CI [1.50, 6.80], $p=.003$ (Table~\ref{tab:robustness}).

\subsection{Conversational style and structure}

We examined whether sensing changed the chatbot’s response style or the structure of the conversation.
Reply length did not differ reliably between conditions (mean sensing-on difference: 9.0 non-space characters, $p=.28$).
However, an average of 67.9\% of sensing-on replies per conversation contained a question mark, compared with 91.9\% of sensing-off replies (difference: 24.0 percentage points, $p=.001$). Sensing-on conversations also averaged 0.77 fewer turns ($p=.010$).
Together, these differences suggest that sensing shifted the interaction toward a more declarative style, with less frequent questioning potentially contributing to shorter exchanges. Whether this shift contributed to the higher persuasiveness ratings remains an open question.

Selected replies illustrate both this aggregate pattern and its limits. For example, a sensing-on reply to P10 made a direct assertion: \enquote{Relying on AI tools to collect data and reference material weakens students' independent learning and critical thinking because they no longer need to investigate and evaluate information themselves.} By contrast, a sensing-off reply to P4 used two rhetorical questions: \enquote{Is the value of art limited to human creation? If AI-generated art can also move people and provoke thought, why should it not be valued equally?}
Sensing-on replies could nevertheless be interrogative. In response to P9, the chatbot asked: \enquote{If we begin relying on AI at any stage of writing, are we not gradually weakening writers' ability to engage with complex emotions and ideas, ultimately risking the individuality and depth of their work?} All three excerpts are translated from Mandarin.
Together with the aggregate results, these examples suggest that sensing-on replies were more often declarative, but both conditions produced a range of rhetorical forms.

\subsection{Affect}

Negative affect was lower after disclosure, $F(1,23)=8.97$, $p=.006$, $\eta_p^2=.28$ ($M=13.15$ vs. $15.50$), and slightly higher with sensing, $F(1,23)=4.26$, $p=.050$, $\eta_p^2=.16$ (Fig.~\ref{fig:results}, right). There was no interaction, $p=.72$. No statistically significant main effects or interactions were found for positive affect or SAM valence, arousal, and dominance. 

\subsection{Reading disclosure together with session order}
\label{sec:order}

\begin{figure}[t]
  \centering
  \includegraphics[width=\linewidth]{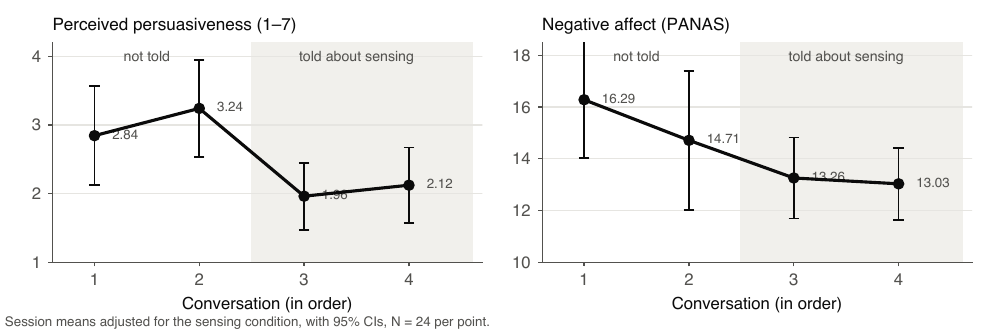}
  \caption{Perceived persuasiveness and negative affect by conversation order, adjusted for sensing. Persuasiveness shows a discrete drop at the disclosure boundary; negative affect declines more steadily across the session.}
  \Description{Two line plots show adjusted means over four conversations. Persuasiveness rises slightly from conversation one to two, drops after disclosure, then remains similar. Negative affect declines from conversation one through three.}
  \label{fig:sessions}
\end{figure}

The post-disclosure blocks always followed the pre-disclosure blocks, so awareness was confounded with session order. To examine whether the between-block difference reflected a gradual change over time, we modeled a linear trend across the four conversations together with a separate change at the disclosure boundary.
For perceived persuasiveness, the linear trend was not significant ($b=+0.28$, $p=.25$), whereas the model estimated a 1.57-point drop at the disclosure boundary ($b=-1.57$, $p=.006$). Sensing-adjusted means increased from 2.84 to 3.24 before disclosure, dropped to 1.96 immediately afterward, and then remained similar at 2.12 (Fig.~\ref{fig:sessions}).
Ratings therefore showed a midpoint post-disclosure drop rather than a steady linear decline, although we cannot attribute this drop to disclosure alone.
Negative affect followed a different pattern: it declined gradually across the session, with no additional change at the disclosure boundary ($p=.68$), suggessting that habituation is more plausible than a disclosure effect.

\subsection{How participants described the interaction}

Interview excerpts shows there is a gap between the persuasiveness ratings and the actual opinion movement. Some participants reported that the chatbot's arguments were valid, but that they were already committed to their position. Others described the conversation as persuasive, but did not report a change in their attitude.
P13 said, \enquote{It didn't change my mind, but that's only because I already had a very strong opinion. I actually thought the points it made were quite valid.} Sensing also remained hard to recognize after disclosure. P3 explained, \enquote{I knew it was using my biosignal, but during the chat, I couldn't really tell what it was doing with it.}

\section{Discussion}

\subsection*{Physiological sensing as dynamic persuasive tailoring}

Prior work on persuasive tailoring has largely adapted messages to relatively stable information about a person prior to the conversation, such as demographics or personality~\cite{Fogg2002PersuasiveTechnology,kaptein2012heterogeneity,kaptein2015personalizing,matz2024potential}.
Physiological sensing introduces a different form of tailoring that changes as the conversation unfolds.
Our findings show that such dynamic adaptation matters.
Before disclosure, the sensing-on chatbot was rated more persuasive than the sensing-off chatbot, including after adjustment for condition order. Across both blocks, sensing was also associated with greater immediate movement toward the position advocated by the chatbot.

Note that the result should not be read as evidence that the chatbot accurately recognized emotion or that every sensed reply was better tailored.
Because the sensing-on condition combined physiological features with instructions to infer a state and adapt the response, these components jointly shape the chatbot's responses.

\subsection*{Sensing shifted conversational form}

The dialogue logs show that sensing changed the chatbot's conversational style.
When sensing was on, chatbot replies contained question marks less often, and conversations ended after fewer turns.
Sensing-on dialogue was therefore more statement-led and declarative, with shorter exchanges.
Persuasion research shows that rhetorical questions can shape how people process an argument~\cite{petty1981rhetorical}, which provides a potential explaination for the higher persuasiveness ratings in the sensing-on condition.

\subsection*{Disclosure changed evaluation more clearly than attitude movement}

Persuasiveness ratings dropped at the disclosure boundary, and the sensing-on advantage was smaller in the post-disclosure block. The Persuasion Knowledge Model provides one interpretation: recognizing an influence tactic gives people new grounds on which to evaluate the persuader~\cite{friestad1994persuasion,campbell2000consumers}. This interpretation is also consistent with studies in which AI disclosure changed evaluations more readily than behavioural responses~\cite{Lim2023Disclosure,Bashardoust2024ShareGenAI,bohm2023devalueLLM}. In our study, attitude movement did not show a corresponding block difference.

Note that this pattern does not establish that disclosure activated persuasion knowledge or reduced persuasiveness. We did not measure persuasion knowledge, and disclosure always occurred at the session midpoint. The results show only that evaluations changed sharply in the post-disclosure block, whereas attitude direction did not.
A between-participants manipulation of disclosure is needed to separate awareness from session order and to test whether persuasion knowledge accounts for this difference.

\section{Design implications: from disclosure to inspectable adaptation}

Physiological data can support different purposes, from operating a device and supporting wellbeing to adapting persuasive messages. Our study highlights why these uses should be distinguished: knowing that a device collects biosignals does not necessarily make it apparent that those signals are shaping an argument. This gap between disclosure and understanding motivates three design considerations for making physiological adaptation more inspectable.

\subsection*{Show when sensed data are in use.} Systems should provide a persistent indication when physiological data are available to an agent and make clear when they are used to adapt generated responses.
This indication should describe the data flow rather than simply present an uncertain state inference. Layered explanations could provide more detail on demand without interrupting every conversational turn~\cite{lim2009assessing,nahum2016just}.

\subsection*{Provide a genuine non-sensed mode.} Users should be able to disable physiological adaptation without losing access to the provided service.
Disabling it should remove both the physiological features and the instructions that act on them. Controls should remain accessible during the interaction, rather than appearing only in an initial consent or disclosure screen.

\subsection*{Maintain an audit trail.} System logs should record which physiological signals were available, when they were provided to the model, and which adaptation instructions were active. This record would support independent evaluation of how sensed data were used during the interaction.

\section{Limitations}
\label{sec:limitations}

Our study has four main limitations. First, awareness was fixed to the second half of the session, so disclosure effects cannot be separated from session order. Second, each condition used a different set of self-selected topics, which may have introduced topic-related variation. Third, the sample was small ($N=24$) and university-based, limiting the generalizability of the findings. Finally, we evaluated only on GPT-4 Turbo, so the results may not generalize to other LLMs.

\section{Conclusion}

This paper examined whether physiological signals from a VR headset change the persuasiveness of an LLM-based chatbot, both before and after sensing was disclosed.
Before disclosure, participants rated the chatbot as more persuasive when sensing was on. Across the study, sensing was also associated with greater immediate movement toward the chatbot’s position.
Dialogue logs showed a more declarative style when sensing was on, with fewer replies containing question marks and fewer conversational turns.
Persuasiveness ratings were lower after disclosure, although this difference cannot be separated from session order.
These findings motivate making biosignal-driven adaptation inspectable and providing a genuine non-sensed mode.

\bibliographystyle{ACM-Reference-Format}
\bibliography{vr_ref}

@article{Bashardoust2024ShareGenAI,
  author = {Bashardoust, Amirsiavosh and Feuerriegel, Stefan and Shrestha, Yash Raj},
  title = {Comparing the Willingness to Share for Human-generated vs. AI-generated Fake News},
  year = {2024},
  issue_date = {November 2024},
  publisher = {Association for Computing Machinery},
  address = {New York, NY, USA},
  volume = {8},
  number = {CSCW2},
  url = {https://doi.org/10.1145/3687028},
  doi = {10.1145/3687028},
  journal = {Proc. ACM Hum.-Comput. Interact.},
  month = nov,
  articleno = {489},
  numpages = {21}
}

@article{costello2024reduceConspiracy,
  title={Durably reducing conspiracy beliefs through dialogues with AI},
  author={Costello, Thomas H and Pennycook, Gordon and Rand, David G},
  journal={Science},
  volume={385},
  number={6714},
  pages={eadq1814},
  year={2024},
  publisher={American Association for the Advancement of Science}
}

@misc{bohm2023devalueLLM,
  title={People devalue generative AI’s competence but not its advice in addressing societal and personal challenges. Communications Psychology 1, 1 (Nov. 2023), 1--10},
  author={B{\"o}hm, Robert and J{\"o}rling, Moritz and Reiter, Leonhard and Fuchs, Christoph},
  year={2023}
}

@article{palmer2023governance,
  title={Large language models can argue in convincing ways about politics, but humans dislike AI authors: implications for governance},
  author={Palmer, Alexis and Spirling, Arthur},
  journal={Political science},
  volume={75},
  number={3},
  pages={281--291},
  year={2023},
  publisher={Taylor \& Francis}
}

@ARTICLE{Lim2023Disclosure,
  title         = "The effect of source disclosure on evaluation of
                   {AI}-generated messages: A two-part study",
  author        = "Lim, Sue and Schmälzle, Ralf",
  journal       = "arXiv [cs.CL]",
  month         =  nov,
  year          =  2023,
  archivePrefix = "arXiv",
  primaryClass  = "cs.CL"
}

@inproceedings{zanbaka2006cat,
  title={Can a virtual cat persuade you? The role of gender and realism in speaker persuasiveness},
  author={Zanbaka, Catherine and Goolkasian, Paula and Hodges, Larry},
  booktitle={Proceedings of the SIGCHI conference on Human Factors in computing systems},
  pages={1153--1162},
  year={2006}
}

@article{compton2013inoculation,
  title={Inoculation theory},
  author={Compton, Josh},
  journal={The SAGE handbook of persuasion: Developments in theory and practice},
  volume={2},
  pages={220--237},
  year={2013},
  publisher={Thousand Oaks}
}

@ARTICLE{Salvi2025Persuade,
  title     = "On the conversational persuasiveness of {GPT}-4",
  author    = "Salvi, Francesco and Horta Ribeiro, Manoel and Gallotti, Riccardo
               and West, Robert",
  journal   = "Nat. Hum. Behav.",
  publisher = "Springer Science and Business Media LLC",
  volume    =  9,
  number    =  8,
  pages     = "1645--1653",
  month     =  aug,
  year      =  2025,
  language  = "en"
}

@article{idrees2024persuasive,
  title={Persuasive technologies design for mental and behavioral health platforms: A scoping literature review},
  author={Idrees, Abdul Rahman and Kraft, Robin and Mutter, Agnes and Baumeister, Harald and Reichert, Manfred and Pryss, R{\"u}diger},
  journal={PLOS Digital Health},
  volume={3},
  number={5},
  pages={e0000498},
  year={2024},
  publisher={Public Library of Science San Francisco, CA USA}
}

@article{salehzadeh2024wearable,
  title={Wearable activity trackers: A survey on utility, privacy, and security},
  author={Salehzadeh Niksirat, Kavous and Velykoivanenko, Lev and Zufferey, No{\'e} and Cherubini, Mauro and Huguenin, K{\'e}vin and Humbert, Mathias},
  journal={ACM Computing Surveys},
  volume={56},
  number={7},
  pages={1--40},
  year={2024},
  publisher={ACM New York, NY}
}

@inproceedings{neupane2025wearable,
  title={Wearable Meets LLM for Stress Management: A Duoethnographic Study Integrating Wearable-Triggered Stressors and LLM Chatbots for Personalized Interventions},
  author={Neupane, Sameer and Dongre, Poorvesh and Gracanin, Denis and Kumar, Santosh},
  booktitle={Proceedings of the Extended Abstracts of the CHI Conference on Human Factors in Computing Systems},
  year={2025},
  publisher={Association for Computing Machinery},
  address={New York, NY, USA},
  articleno={588},
  numpages={8},
  doi={10.1145/3706599.3720197},
  series={CHI EA '25}
}

@ARTICLE{Bai2025LLMPersuation,
  title    = "{LLM}-generated messages can persuade humans on policy issues",
  author   = "Bai, Hui and Voelkel, Jan G and Muldowney, Shane and Eichstaedt,
              Johannes C and Willer, Robb",
  journal  = "Nat. Commun.",
  volume   =  16,
  number   =  1,
  pages    =  6037,
  month    =  jul,
  year     =  2025,
  language = "en"
}

@article{matz2024potential,
  title={The potential of generative AI for personalized persuasion at scale},
  author={Matz, Sandra C and Teeny, Jacob D and Vaid, Sumer S and Peters, Heinrich and Harari, Gabriella M and Cerf, Moran},
  journal={Scientific Reports},
  volume={14},
  number={1},
  pages={4692},
  year={2024},
  publisher={Nature Publishing Group UK London}
}

@ARTICLE{Spelt2022Physiology,
  title     = "Physiology-based personalization of persuasive technology: a user
               modeling perspective",
  author    = "Spelt, Hanne A A and Westerink, Joyce H D M and Frank, Lily and
               Ham, Jaap and IJsselsteijn, Wijnand A",
  journal   = "User Model. User-adapt Interact.",
  publisher = "Springer Science and Business Media LLC",
  volume    =  32,
  number    = "1-2",
  pages     = "133--163",
  month     =  apr,
  year      =  2022,
  language  = "en"
}

@article{danry2024deceptive,
  title={Deceptive AI systems that give explanations are more convincing than honest AI systems and can amplify belief in misinformation},
  author={Danry, Valdemar and Pataranutaporn, Pat and Groh, Matthew and Epstein, Ziv and Maes, Pattie},
  journal={arXiv preprint arXiv:2408.00024},
  year={2024}
}

@article{liu2025llm,
  title={LLM can be a dangerous persuader: Empirical study of persuasion safety in large language models},
  author={Liu, Minqian and Xu, Zhiyang and Zhang, Xinyi and An, Heajun and Qadir, Sarvech and Zhang, Qi and Wisniewski, Pamela J and Cho, Jin-Hee and Lee, Sang Won and Jia, Ruoxi and others},
  journal={arXiv preprint arXiv:2504.10430},
  year={2025}
}

@article{Fogg2002PersuasiveTechnology,
author = {Fogg, B. J.},
title = {Persuasive technology: using computers to change what we think and do},
year = {2002},
issue_date = {December 1 - December 31, 2002},
publisher = {Association for Computing Machinery},
address = {New York, NY, USA},
volume = {2002},
number = {December},
url = {https://doi.org/10.1145/764008.763957},
doi = {10.1145/764008.763957},
journal = {Ubiquity},
month = dec,
articleno = {5},
numpages = {32}
}

@article{kaptein2012heterogeneity,
  title={Heterogeneity in the effects of online persuasion},
  author={Kaptein, Maurits and Eckles, Dean},
  journal={Journal of Interactive Marketing},
  volume={26},
  number={3},
  pages={176--188},
  year={2012},
  publisher={SAGE Publications Sage CA: Los Angeles, CA}
}

@inproceedings{fang2024physiollm,
  title={Physiollm: Supporting personalized health insights with wearables and large language models},
  author={Fang, Cathy Mengying and Danry, Valdemar and Whitmore, Nathan and Bao, Andria and Hutchison, Andrew and Pierce, Cayden and Maes, Pattie},
  booktitle={2024 IEEE EMBS International Conference on Biomedical and Health Informatics (BHI)},
  pages={1--8},
  year={2024},
  organization={IEEE}
}

@inproceedings{dongre2024physiology,
  title={Physiology-driven empathic large language models (EmLLMs) for mental health support},
  author={Dongre, Poorvesh},
  booktitle={Extended Abstracts of the CHI Conference on Human Factors in Computing Systems},
  pages={1--5},
  year={2024}
}

@article{kaptein2015personalizing,
  title={Personalizing persuasive technologies: Explicit and implicit personalization using persuasion profiles},
  author={Kaptein, Maurits and Markopoulos, Panos and De Ruyter, Boris and Aarts, Emile},
  journal={International Journal of Human-Computer Studies},
  volume={77},
  pages={38--51},
  year={2015},
  publisher={Elsevier}
}

@inproceedings{schmidt2018introducing,
  title={Introducing wesad, a multimodal dataset for wearable stress and affect detection},
  author={Schmidt, Philip and Reiss, Attila and Duerichen, Robert and Marberger, Claus and Van Laerhoven, Kristof},
  booktitle={Proceedings of the 20th ACM international conference on multimodal interaction},
  pages={400--408},
  year={2018}
}

@article{liu2023biosignal,
  title={BioSignal Copilot: Leveraging the power of LLMs in drafting reports for biomedical signals},
  author={Liu, Chunyu and Ma, Yongpei and Kothur, Kavitha and Nikpour, Armin and Kavehei, Omid},
  journal={medRxiv},
  pages={2023--06},
  year={2023},
  publisher={Cold Spring Harbor Laboratory Press}
}

@inproceedings{wu2024mindshift,
  title={MindShift: leveraging large language models for mental-states-based problematic smartphone use intervention},
  author={Wu, Ruolan and Yu, Chun and Pan, Xiaole and Liu, Yujia and Zhang, Ningning and Fu, Yue and Wang, Yuhan and Zheng, Zhi and Chen, Li and Jiang, Qiaolei and others},
  booktitle={Proceedings of the 2024 CHI conference on human factors in computing systems},
  pages={1--24},
  year={2024}
}

@inproceedings{Genc2025Persuasion,
author = {Gen\c{c}, H\"{u}seyin U\u{g}ur and Chandrasegaran, Senthil and Dingler, Tilman and Verma, Himanshu},
title = {Persuasion in Pixels and Prose: The Effects of Emotional Language and Visuals in Agent Conversations on Decision-Making},
year = {2025},
isbn = {9798400713941},
publisher = {Association for Computing Machinery},
address = {New York, NY, USA},
url = {https://doi.org/10.1145/3706598.3713579},
doi = {10.1145/3706598.3713579},
booktitle = {Proceedings of the 2025 CHI Conference on Human Factors in Computing Systems},
articleno = {252},
numpages = {27},
location = {
},
series = {CHI '25}
}

@article{bradley1994measuring,
  title={Measuring emotion: the self-assessment manikin and the semantic differential},
  author={Bradley, Margaret M and Lang, Peter J},
  journal={Journal of behavior therapy and experimental psychiatry},
  volume={25},
  number={1},
  pages={49--59},
  year={1994},
  publisher={Elsevier}
}

@article{watson1988development,
  title={Development and validation of brief measures of positive and negative affect: the PANAS scales.},
  author={Watson, David and Clark, Lee Anna and Tellegen, Auke},
  journal={Journal of personality and social psychology},
  volume={54},
  number={6},
  pages={1063},
  year={1988},
  publisher={American Psychological Association}
}

@article{iriarte2021evaluating,
  title={Evaluating user experience with physiological monitoring: A systematic literature review},
  author={Iriarte, Ainhoa Apraiz and Erle, Ganix Lasa and Etxabe, Maitane Mazmela},
  journal={DYNA New Technologies Journal},
  volume={8},
  number={1},
  year={2021}
}

@inproceedings{grubert2018text,
  title={Text entry in immersive head-mounted display-based virtual reality using standard keyboards},
  author={Grubert, Jens and Witzani, Lukas and Ofek, Eyal and Pahud, Michel and Kranz, Matthias and Kristensson, Per Ola},
  booktitle={2018 IEEE Conference on Virtual Reality and 3D User Interfaces (VR)},
  pages={159--166},
  year={2018},
  organization={IEEE}
}

@inproceedings{dube2019text,
  title={Text entry in virtual reality: A comprehensive review of the literature},
  author={Dube, Tafadzwa Joseph and Arif, Ahmed Sabbir},
  booktitle={International Conference on Human-Computer Interaction},
  pages={419--437},
  year={2019},
  organization={Springer}
}

@inproceedings{heidicker2017influence,
  title={Influence of avatar appearance on presence in social VR},
  author={Heidicker, Paul and Langbehn, Eike and Steinicke, Frank},
  booktitle={2017 IEEE symposium on 3D user interfaces (3DUI)},
  pages={233--234},
  year={2017},
  organization={IEEE}
}

@article{poulton1982influential,
  title={Influential companions: Effects of one strategy on another in the within-subjects designs of cognitive psychology.},
  author={Poulton, EC},
  journal={Psychological Bulletin},
  volume={91},
  number={3},
  pages={673},
  year={1982},
  publisher={American Psychological Association}
}

@article{kreibig2010autonomic,
  title={Autonomic nervous system activity in emotion: A review},
  author={Kreibig, Sylvia D},
  journal={Biological psychology},
  volume={84},
  number={3},
  pages={394--421},
  year={2010},
  publisher={Elsevier}
}

@article{levenson2014autonomic,
  title={The autonomic nervous system and emotion},
  author={Levenson, Robert W},
  journal={Emotion review},
  volume={6},
  number={2},
  pages={100--112},
  year={2014},
  publisher={Sage Publications Sage UK: London, England}
}

@article{nahum2016just,
  title={Just-in-time adaptive interventions (JITAIs) in mobile health: key components and design principles for ongoing health behavior support},
  author={Nahum-Shani, Inbal and Smith, Shawna N and Spring, Bonnie J and Collins, Linda M and Witkiewitz, Katie and Tewari, Ambuj and Murphy, Susan A},
  journal={Annals of behavioral medicine},
  pages={1--17},
  year={2016},
  publisher={Springer}
}

@inproceedings{lim2009assessing,
  title={Assessing demand for intelligibility in context-aware applications},
  author={Lim, Brian Y and Dey, Anind K},
  booktitle={Proceedings of the 11th international conference on Ubiquitous computing},
  pages={195--204},
  year={2009}
}

@article{doi:10.1073/pnas.2403116121,
    author = {Kobi Hackenburg  and Helen Margetts },
    title = {Evaluating the persuasive influence of political microtargeting with large language models},
    journal = {Proceedings of the National Academy of Sciences},
    volume = {121},
    number = {24},
    pages = {e2403116121},
    year = {2024},
    doi = {10.1073/pnas.2403116121},
    URL = {https://www.pnas.org/doi/abs/10.1073/pnas.2403116121},
    eprint = {https://www.pnas.org/doi/pdf/10.1073/pnas.2403116121}
}

@article{hackenburg2025levers,
  title={The levers of political persuasion with conversational artificial intelligence},
  author={Hackenburg, Kobi and Tappin, Ben M and Hewitt, Luke and Saunders, Ed and Black, Sid and Lin, Hause and Fist, Catherine and Margetts, Helen and Rand, David G and Summerfield, Christopher},
  journal={Science},
  volume={390},
  number={6777},
  pages={eaea3884},
  year={2025},
  publisher={American Association for the Advancement of Science}
}

@article{friestad1994persuasion,
  title={The Persuasion Knowledge Model: How People Cope with Persuasion Attempts},
  author={Friestad, Marian and Wright, Peter},
  journal={Journal of Consumer Research},
  volume={21},
  number={1},
  pages={1--31},
  year={1994},
  publisher={Oxford University Press},
  doi={10.1086/209380}
}

@article{campbell2000consumers,
  title={Consumers' Use of Persuasion Knowledge: The Effects of Accessibility and Cognitive Capacity on Perceptions of an Influence Agent},
  author={Campbell, Margaret C. and Kirmani, Amna},
  journal={Journal of Consumer Research},
  volume={27},
  number={1},
  pages={69--83},
  year={2000},
  publisher={Oxford University Press},
  doi={10.1086/314309}
}

@article{petty1981rhetorical,
  title={Effects of Rhetorical Questions on Persuasion: A Cognitive Response Analysis},
  author={Petty, Richard E. and Cacioppo, John T. and Heesacker, Martin},
  journal={Journal of Personality and Social Psychology},
  volume={40},
  number={3},
  pages={432--440},
  year={1981},
  publisher={American Psychological Association},
  doi={10.1037/0022-3514.40.3.432}
}

@article{petty1977forewarning,
  title={Forewarning, Cognitive Responding, and Resistance to Persuasion},
  author={Petty, Richard E. and Cacioppo, John T.},
  journal={Journal of Personality and Social Psychology},
  volume={35},
  number={9},
  pages={645--655},
  year={1977},
  publisher={American Psychological Association},
  doi={10.1037/0022-3514.35.9.645}
}

@article{mccambridge2014hawthorne,
  title={Systematic Review of the {Hawthorne} Effect: New Concepts Are Needed to Study Research Participation Effects},
  author={McCambridge, Jim and Witton, John and Elbourne, Diana R.},
  journal={Journal of Clinical Epidemiology},
  volume={67},
  number={3},
  pages={267--277},
  year={2014},
  publisher={Elsevier},
  doi={10.1016/j.jclinepi.2013.08.015}
}

@inproceedings{genc2026bots,
  author={Gen\c{c}, H\"{u}seyin U\u{g}ur and Gu, Heng and Degachi, Chadha and Niforatos, Evangelos and Chandrasegaran, Senthil and Verma, Himanshu},
  title={The Bots of Persuasion: Examining How Conversational Agents' Linguistic Expressions of Personality Affect User Perceptions and Decisions},
  booktitle={Proceedings of the 2026 CHI Conference on Human Factors in Computing Systems},
  year={2026},
  publisher={Association for Computing Machinery},
  address={New York, NY, USA},
  articleno={796},
  numpages={25},
  doi={10.1145/3772318.3791407},
  series={CHI '26}
}

@inproceedings{yeo2026deception,
  author={Yeo, Haein and Jin, Seungwan and Noh, Taehyung and Shin, Yejin and Kang, Sangyeon and Heo, Sangwoo and Chung, Jiwon and Hyun, Hwarim and Han, Kyungsik},
  title={{Can LLMs Persuade Humans with Deception?}: From a Deceptive Strategy Taxonomy to a Large-Scale Empirical Study},
  booktitle={Proceedings of the 2026 CHI Conference on Human Factors in Computing Systems},
  year={2026},
  publisher={Association for Computing Machinery},
  address={New York, NY, USA},
  articleno={1506},
  numpages={21},
  doi={10.1145/3772318.3791188},
  series={CHI '26}
}

@inproceedings{wang2025healthguru,
  author={Wang, Xingbo and Griffith, Janessa and Adler, Daniel A. and Castillo, Joey and Choudhury, Tanzeem and Wang, Fei},
  title={Exploring Personalized Health Support through Data-Driven, Theory-Guided {LLM}s: A Case Study in Sleep Health},
  booktitle={Proceedings of the 2025 CHI Conference on Human Factors in Computing Systems},
  year={2025},
  publisher={Association for Computing Machinery},
  address={New York, NY, USA},
  numpages={15},
  doi={10.1145/3706598.3713852},
  series={CHI '25}
}

@inproceedings{liu2026sensing,
  author={Liu, Ailin and Karoui, Yesmine and Draxler, Fiona and Kreuter, Frauke and Chiossi, Francesco},
  title={Sensing What Surveys Miss: Understanding and Personalizing Proactive {LLM} Support by User Modeling},
  booktitle={Proceedings of the 2026 CHI Conference on Human Factors in Computing Systems},
  year={2026},
  publisher={Association for Computing Machinery},
  address={New York, NY, USA},
  articleno={1012},
  numpages={21},
  doi={10.1145/3772318.3791191},
  series={CHI '26}
}

@inproceedings{gamage2025labeling,
  author={Gamage, Dilrukshi and Sewwandi, Dilki and Zhang, Min and Bandara, Arosha K.},
  title={Labeling Synthetic Content: User Perceptions of Label Designs for {AI}-Generated Content on Social Media},
  booktitle={Proceedings of the 2025 CHI Conference on Human Factors in Computing Systems},
  year={2025},
  publisher={Association for Computing Machinery},
  address={New York, NY, USA},
  articleno={814},
  numpages={20},
  doi={10.1145/3706598.3713171},
  series={CHI '25}
}

@inproceedings{khatiwada2026news,
  author={Khatiwada, Prerana and Pappu, Varun and Bagozzi, Benjamin E. and Mauriello, Matthew Louis},
  title={When {AI} Rewrites the News: How Sentiment, Framing, and {LLM} Disclosure Shape Perceptions},
  booktitle={Proceedings of the 2026 CHI Conference on Human Factors in Computing Systems},
  year={2026},
  publisher={Association for Computing Machinery},
  address={New York, NY, USA},
  articleno={1241},
  numpages={25},
  doi={10.1145/3772318.3791527},
  series={CHI '26}
}

\end{document}